\documentclass[aps,prl,twocolumn,groupedaddress,amsmath,amssymb]{revtex4}
\usepackage{graphicx}
\usepackage{bm}
\usepackage{multirow,amssymb,amsbsy,amsmath}

\newcommand{\Funktion}[2]{#1\kern-0.2em\left(#2\right)}

\newcommand{\trtxt}[2][]{\text{Tr}_{#1}\{#2\}}
\newcommand*{\bra}[1]{\mathopen{\langle}#1\mathclose{|}}
\newcommand*{\ket}[1]{\mathopen{|}#1\mathclose{\rangle}}

\newcommand{\refeq}[1]{Eq.~(\ref{#1})}

\newcommand{\comu}[2]{\left[#1,#2\right]}

\begin{document}

\preprint{APS/123-QED}

\title{Exploiting Quench Dynamics in Spin Chains for Distant Entanglement and Quantum Communication}

\author{Hannu Wichterich}
\affiliation{
    Dept.~of Physics and Astronomy, %
    University College London, %
    Gower Street, %
    WC1E 6BT London, %
    United Kingdom}
\author{Sougato Bose}
\affiliation{
    Dept.~of Physics and Astronomy, %
    University College London, %
    Gower Street, %
    WC1E 6BT London, %
    United Kingdom}
\date{\today}

\begin{abstract}
We suggest a method of entangling significantly the distant ends of
a spin chain using minimal control. This entanglement between
distant individual spins is brought about solely by exploiting the
dynamics of an initial mixed state with N\'eel order if the lattice
features nearest-neighbor XXZ interaction. There is no need to
control single spins or to have engineered couplings or to pulse
globally. The method only requires an initial non-adiabatic switch
(a quench) between two Hamiltonians followed by evolution under the
second Hamiltonian. The scheme is robust to randomness of the
couplings as well as the finiteness of an appropriate quench and
could potentially be implemented in various experimental setups,
ranging from atoms in optical lattices to Josephson junction arrays.
\end{abstract}


\maketitle

The objective of quantum communication is to transfer a quantum
state $\ket{\phi}$ from a sender (Alice) to a receiver (Bob) as
accurately as possible. For this, Alice can simply encode
$\ket{\phi}$ on a carrier and send it. Alternatively, she and Bob
can use teleportation \cite{Bennett1993}, for which they need to
share a pair of particles in an entangled state
$\ket{\psi^{+}}=\frac{1}{\sqrt{2}}(\ket{\downarrow,
\uparrow}+\ket{\uparrow, \downarrow})$. $\ket{\psi^{+}}$, with
appropriate operations and classical communications,  enables the
noiseless transmission of a state $\ket{\phi}$ from Alice to Bob.
Photons are ideal for long distance quantum communication.

This paper is based on a manifestation of quantum communication
which aims to connect distinct parts, i.e. registers, of a quantum
computer. The idea is to use a many-body system (with permanent
interactions) incorporating the sender, the receiver, and the
mediating channel all together. All components are then stationary,
and could be made of the same units. This setup does not require the
interfacing of stationary qubits with photons. This could
considerably reduce the complexity of interconnects between
registers of a quantum computer which typically need to enable
entanglement sharing over short distances i.e. several lattice
sites. For the same reason, the physical movement of ions is
considered seriously for communication between ion trap quantum
registers \cite{Kielpinski2002}.

\begin{figure}
 \includegraphics[width=0.48\textwidth]{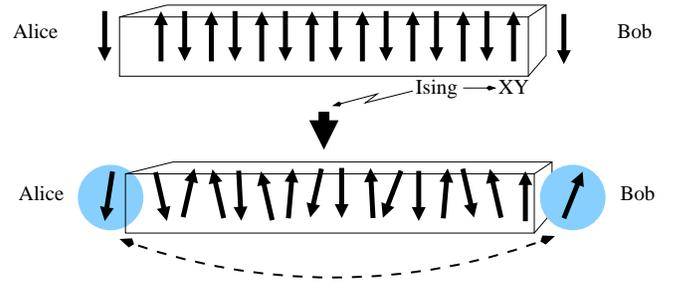}
\caption{(Color online) Schematic of our proposal of entangling distant spins.
Alice and Bob are at opposite ends of the chain. First, the spin
chain is initialized in the antiferromagnetic Ising groundstate. By
a non-adiabatic switching to an XY interaction and subsequent time
evolution quantum correlations (entanglement) is being established
between Alice's and Bob's spins. \label{fig.pic}}
\end{figure}

Fig.~\ref{fig.pic} pictures our scenario, where Alice and Bob are
situated at opposite ends of a one dimensional (1D) lattice of
perpetually interacting spin-$\frac12$ particles. We suggest a
scheme which allows the establishment of a strong entanglement
between Alice's and Bob's spins (the remotest spins of the lattice)
without any requirement of local control for initializing the chain
or for the subsequent dynamics. Neither the repeated switchings of
any fields (local or global) nor any specially engineered couplings
are required. This is close to the scenario of much work on state
transfer through spin chains (e.g. Refs.\cite{Bose2003, Osborne2004,
Christandl2004, Fitzsimons2006, Giovannetti2006}), but here the
lattice interactions ``generate" entanglement, as opposed to just
``transferring" it. State transfer can itself be modified to yield
entanglement generation schemes (eg. see \cite{D'Amico2007}), but
without a price (e.g. engineered couplings and local preparation),
the amount of entanglement will be very small. Even without such
price, our current mechanism provides a very high entanglement.

In our scheme, first the lattice of strongly interacting spins is
cooled to its ground state. Then, upon instantly changing a global
parameter in its Hamiltonian ({\em i.e.}, performing a {\em quench}
\cite{Sengupta2004,Cramer2008,Calabrese2005, DeChiara2006,
Cincio2007}), the pair of edge spins evolve to a highly entangled
mixed state. This state is special in the sense that entanglement
purification \cite{Bennett1996}, a process that Alice and Bob can
use to convert, by local actions, a few (say $n$) copies of the
state to $m<n$ pure $\ket{\psi^{+}}$ states, can be quite efficient.
These $\ket{\psi^{+}}$ could then be used to teleport any state from
Alice to Bob. Though the scheme of this paper has a qualitative
similarity with entangling the ends of a chain of uncoupled systems
by a sudden switching of interactions
\cite{Eisert2004,Tsomokos2007}, it yields a much higher entanglement
scaling better with the length of the chain. Also, as we will show,
we actually start from a mixed initial state despite the cooling --
so the production of a high entanglement between ends is quite
interesting.

Quenches have been actively studied in condensed matter, usually in
the thermodynamic limit \cite{Sengupta2004} and were recently shown
shown to make local subsystems relax towards maximum entropy
\cite{Cramer2008}. In contrast, here we investigate whether a
quench, when performed on a \textit{finite} system, can produce a
``long-range" entanglement {\em between individual distant spins}.
Recently, the development of a {\em qualitatively different} form of
entanglement, namely that between complementary blocks of a spin
chain was shown to be possible after a quench \cite{Calabrese2005,
DeChiara2006, Cincio2007}. This phenomenon was attributed to a
heuristic picture in which several sources in a lattice emit
oppositely moving pairs of entangled quasiparticles which reach
distinct blocks to entangle them. However, a high entanglement
between two individual spins is not guaranteed by the above picture
-- indeed for certain quenches in the quantum Ising model, where the
above process entangles blocks \cite{Calabrese2005}, there is hardly
any entanglement between the end spins. A more intricate process
involving reflections of wavepackets at boundaries, and a cumulative
contribution of the {\em same} entangled state from different
sources to the end spins, can, for the quench we examine, create a
very high entanglement between end spins at a special time. This is
both directly relevant to quantum communications and {\em
measurable}, as opposed to the entanglement of blocks.

Using a global quench to create entanglement at a distance is highly
attractive because of the non-requirement of controlling single
spins. Other low control ways of creating sizeable entanglement
between distant spins include exploiting certain spin chains
  whose end spins are coupled weakly to the remaining ones \cite{CamposVenuti2006, CamposVenuti2007}. This is a
rare engineered case, as generically the ground states of spin chain
models are known to exhibit only very short ranged entanglement
\cite{Osborne2002}. Another approach is based on localizable
entanglement \cite{Verstraete2004a} where one entangles the end
spins of a bulk system by individual measurements on the other
spins, which can be challenging.

Returning now to the problem of entangling distant spins through a quench, consider a chain of $N$ spin-$\frac{1}{2}$ systems with nearest-neighbor XXZ interaction
\begin{align}\label{eq.hamiltonian}
 H=\sum\limits_{k=1}^{N-1}\frac{J_k}{2}\big(\sigma_k^{\text x}\sigma_{k+1}^{\text x}+\sigma_k^{\text y}\sigma_{k+1}^{\text y}+\Delta\,\sigma_k^{\text z}\sigma_{k+1}^{\text z}\big)
\end{align}
where the parameters $J_k$ and $\Delta$ denote the coupling strengths at bond $k$
and the anisotropy respectively, and $\sigma_k^{\text x}$,
$\sigma_k^{\text y}$ and $\sigma_k^{\text z}$ denote the Pauli
operators acting on the spin at site $k$. We assume $J_k=J>0$ (homogeneous anisotropic coupling) in most parts of this work, unless we study disorder in later parts of this paper: To that end the couping is modulated as $J_k=J\,(1+\delta_k)$ with normally distributed random numbers $\delta_k$ with mean zero and standard deviation $\delta$. \refeq{eq.hamiltonian}, corresponds to
the Ising-Model for $\Delta \rightarrow\infty$, and the isotropic
XY, or XX model for $\Delta=0$. As $\comu{H}{S_{\text z}}=0$, with
$S_{\text z} = \sum_{k=1}^N\sigma_k^{\text z}$, the total
z-magnetization is a constant of motion.


We first formulate the analytic case of time-evolution of the Ising
ground state, under action of the XX Hamiltonian. This corresponds
to an instantaneous, i.e. idealized quench in the anisotropy
parameter $\Delta_1\rightarrow\Delta_2$ with
$\Delta_1\rightarrow\infty$ and $\Delta_2 = 0$ thereby crossing
critical value  $\Delta= 1$, which separates the N\'eel-Ising-phase
from the XY-phase. Later on, in a purely numerical study we will
allow for $1<\Delta_1<\infty$ and $0<\Delta_2\leq 1$~. For
$\Delta\gg J$ the Ising groundstate gets arbitrarily close to the
ideal N\'eel state, which is twofold degenerate in the absence of an
external field. These ideal N\'eel states arise from the perfectly
polarized state $\ket{\Downarrow_N}$ upon flipping every other spin:
$\ket{\mathcal{N}_1}
\equiv\ket{\downarrow_1,\uparrow_2,\downarrow_3,\cdots}$ and
$\ket{\mathcal{N}_2}
\equiv\ket{\uparrow_1,\downarrow_2,\uparrow_3,\cdots}$~. Note that
these two states turn into each other by a spin flip at each place,
i.e. $\ket{\mathcal{N}_1}=(\prod_{k=1}^{N}\sigma^{\text
x}_k)\ket{\mathcal{N}_2}$ and vice versa. In an experiment, the
initial preparation of the N\'eel-Ising-groundstate, will yield, at
low enough temperatures, an equal mixture of both N\'eel orders, and
negligible admixture of higher energy eigenstates. We adopt the
notion of \textit{thermal ground state} from \cite{Osborne2002} for
\begin{align}\label{eq.thermal_groundstate}
 \rho_0=\frac12\Big(\ket{\mathcal{N}_1}\bra{\mathcal{N}_1}+\ket{\mathcal{N}_2}\bra{\mathcal{N}_2}\Big)~,
\end{align}
which exhibits the same symmetries as the Ising Hamiltonian $H\,(\Delta\rightarrow\infty)$, as opposed to each individual, degenerate ground state of the antiferromagnetic Ising-chain.

After diagonalization following a Jordan-Wigner transformation, we
are provided the explicit time dependence of the local Fermi
creation operators $ c^{\dagger}_k \equiv
\big(\prod_{l=1}^{k-1}\,-\sigma^{\text z}_l\big)\,\sigma^{+}_k$,
with $\sigma^+_k\equiv\frac{1}{2} (\sigma_k^{\text
x}+i\,\sigma_k^{\text y})$, which reads
\begin{align}\label{eq.jw_fermion_timedep}
c^{\dagger}_k(t)&=\sum\limits_{l=1}^{N}\,f_{k,l}(t)\,c^{\dagger}_l\\
f_{k,l}(t)&\equiv\frac{2}{N+1}\sum\limits_{m=1}^{N}\,\sin (q_m\, k)\,\sin (q_m\, l)\,e^{-i\,E_m\,t}~,
\end{align}
with $E_m=2J\cos(q_m)$ and $q_m =\frac{\pi\,m}{N+1}$.

We now investigate the dynamics of the long range nonclassical
correlations, i.e. the entanglement between the first and the last
spin of the chain, provided the system is initially prepared in the
global state (\ref{eq.thermal_groundstate}) and assuming $\Delta=0$
for the subsequent time evolution. The state (reduced density
operator) $\rho_{1,N}$ of spins at sites $1$ and $N$ of the chain in
the $\lbrace\ket{\uparrow\,\uparrow },\ket{\uparrow\,\downarrow
},\ket{\downarrow\,\uparrow },\ket{\downarrow\,\downarrow }\rbrace$
basis has only the following non-zero elements
\begin{eqnarray}\label{eq.rho_nonzero}
\langle
\uparrow\uparrow|\rho_{1,N}|\uparrow\uparrow\rangle&=&\langle\downarrow\downarrow|\rho_{1,N}|\downarrow\downarrow\rangle=a,\langle\uparrow\downarrow|\rho_{1,N}|\uparrow\downarrow\rangle=\nonumber\\
\langle\downarrow\uparrow|\rho_{1,N}|\downarrow\uparrow\rangle&=&b,\langle\uparrow\downarrow|\rho_{1,N}|\downarrow\uparrow\rangle=\langle\downarrow\uparrow|\rho_{1,N}|\uparrow\downarrow\rangle=c,\nonumber\\
\label{eq.rhored_corrfun}
\end{eqnarray}
which are entirely expressible in terms of second moments of the
Fermi operators (\ref{eq.jw_fermion_timedep}).  From $a=\langle
\sigma^{+}_1\sigma^{-}_1\sigma^{+}_N\sigma^{-}_N\rangle$ and
$c=\langle\sigma^{-}_1\sigma^{+}_N\rangle$, where
$\langle\,\cdots\,\rangle=\trtxt{\rho(t)\,\cdots}$ and
$\rho(t)=e^{-iHt}\rho_0 e^{iHt}$, we find
\begin{eqnarray}\label{eq.wick}
a &=& \langle c^{\dagger}_1c_1\rangle_1\,\langle c^{\dagger}_Nc_N\rangle_1 - \langle c^{\dagger}_1c_N\rangle_1\,\langle c^{\dagger}_Nc_1\rangle_1 \\\nonumber
&&\qquad- \frac12 \big( \langle c^{\dagger}_1 c_1\rangle_1 +\langle c^{\dagger}_N c_N\rangle_1-1\big) \\\nonumber
c &=& \frac12\,\big((-1)^{M+1}\,\langle c^{\dagger}_Nc_1\rangle_1 + c.c. \big) .
\end{eqnarray}
Here, $\langle\,\cdots\,\rangle_1$ denotes the expectation value in the Schr\" odinger picture with respect to the initial state $\ket{\mathcal{N}_1}$, which facilitates factoring higher moments according to Wick's theorem, and $M$ is the conserved number of spin up states in the dynamical state $\ket{\mathcal{N}_1(t)}$, i.e. $M=N/2$ for even $N$ and $M=(N-1)/2$ for odd $N$~. The remaining element $b$ is determined by the trace constraint $2a+2b=1$~. Analogous formulas for ground and thermal states of the XX spin chain were reported in \cite{CamposVenuti2007}.  
The second moments occurring in (\ref{eq.wick}) can be conveniently evaluated in the Heisenberg picture, where
\begin{eqnarray}
\langle c^{\dagger}_i(t) c_j(t)\rangle = \sum\limits_{k,l=1}^{N}f_{i,k}(t)f^{\ast}_{j,l}(t)\;\langle c^{\dagger}_k(0) c_l(0)\rangle~.
\end{eqnarray}
As $\langle c^{\dagger}_k(0)
c_l(0)\rangle_1=\delta_{k,l}\delta_{k,2m},\; (m=1,2,\ldots,M$), we
get
\begin{eqnarray}\label{eq.symm_pairs}
c =\frac12 (-1)^{M+1}\sum\limits_{m=1}^M f_{N,2m}(t)f^{\ast}_{1,2m}(t)+c.c.~.
\end{eqnarray}
This expression underlines that quantum correlations between the end spins are a superposition of contributions from spin flips originating from every other site $2m$.

We now examine whether a given supply of systems, all described by
the same mixed state $\rho_{1N}$ from Eq.~(\ref{eq.rho_nonzero}),
can be converted to a smaller subset of $\ket{\psi^{+}}$ states
through local actions by Alice and Bob and classical messaging
between them, for subsequent use in teleportation. A general
criterion \cite{Bennett1996} for this procedure, called entanglement
purification, to be possible for mixed state $\rho$ of two qubits is
expressed in terms of the \textit{fully entangled fraction} $f\equiv
\mathrm{max}(\bra{e}\rho\ket{e})$. Here the maximum is taken with
respect to all maximally entangled states $\lbrace\ket{e}\rbrace$.
The criterion reads $f> \frac12$, and is adapted to our particular
problem as $\mathrm{max}(b+c,b-c)> \frac12$ in view of
Eq.~(\ref{eq.rho_nonzero}) under the trace constraint
$a+b=\frac{1}{2}$.

\begin{figure}
\includegraphics[width=0.44\textwidth]{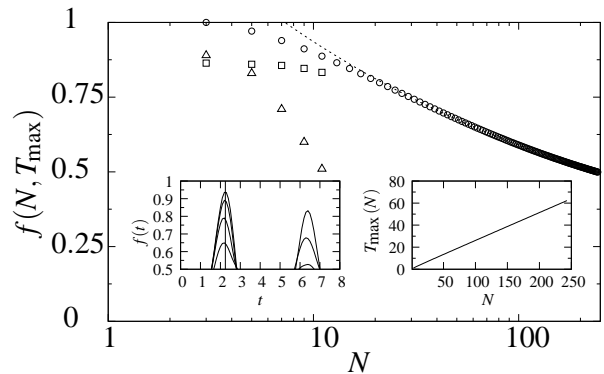}
\caption{Fully entangled fraction $f$ evaluated at time
$t=T_{\text{max}}$ as a function of system size $N$ in a
semilogarithmic scale. The data points for the analytic quench
scenario (circles) exceed $0.5$ for chains up to $N=241$ spins. We
find a good agreement of these data for $N\geq 25$ to a function of
the form $g^{\text{fit}}(N)\propto\,1.42(9)\,N^{-\nu}$~ with
$\nu=0.22(9)$~(dashed curve). For $N\leq 11$ the figure is
supplemented with data from numerical diagonalization for more
general quench characteristics $(\Delta_1\rightarrow\Delta_2)$~:
$(\infty\rightarrow 1)$ (triangles) and $(3\rightarrow 0)$
(squares)~. {\it Left inset}: Typical time evolution of $f$ in a
quench $(\infty\rightarrow 0)$ for $N=7$ with first maximum at
$T_\text{max}$ (vertical line). Disorder ($\delta=0,0.1,0.2,0.3$ from top to bottom
curves respectively) is seen not to affect the amount of
entanglement seriously and leaves $T_{\text{max}}$ almost unchanged
(average taken over 100 realizations). {\it Right inset} : Linear
scaling of $T_{\text{max}}$ with $N$. \label{fig.2}}
\end{figure}

The function $f$ will feature a first
maximum some time $T_{\text{max}}$ after the instant of quench (Fig.\ref{fig.2}, left inset) and
this is always found (by a numerical search for different $N$) to
scale as $T_{\text{max}}\sim\frac{N}{\pi J}$ (Fig.~\ref{fig.2},
right inset -- in fact, $T_{\text{max}}$ is slightly
lower than $\frac{N}{\pi J}$). 
Interestingly, for chains with even $N$ and $\rho_0$ as the initial
state, $\rho_{1,N}$ is always unentangled (separable). This is
because $\ket{\mathcal{N}_1}$ contributes
$|\xi\rangle=\alpha|\downarrow\uparrow\rangle+i\beta|\uparrow\downarrow\rangle$
to $\rho_{1,N}$ (with real $\alpha$ and $\beta$) while symmetry
implies $\ket{\mathcal{N}_2}$ to contribute an equal amount of
$|\tilde{\xi}\rangle=\sigma_x\otimes\sigma_x|\xi\rangle$ so that
$|\xi\rangle\langle\xi|+|\tilde{\xi}\rangle\langle\tilde{\xi}|$ is
separable. Therefore, in Fig.~\ref{fig.2} and henceforth, we
consider {\em odd} $N$.

A high
 entanglement specifically between the end spins requires the
reflection of wavepackets at the boundaries, the initial N\'{e}el
order and an independent propagation of different fermions after
quench. Consider $\ket{\mathcal{N}_1}$ (with odd $N$) to have its
central spin $\ket{\uparrow}$. Each $\ket{\uparrow}$ of
$\ket{\mathcal{N}_1}$ (or each $\ket{\downarrow}$ of
$\ket{\mathcal{N}_2}$) is the location of a fermion which goes to a
{\em superposition} of left and right moving wavepackets due to the
dynamics. From Ref.\cite{Calabrese2005,DeChiara2006} we know that
such dynamics alone suffices to entangle blocks, but entangling
individual spins requires much more. $T_{max}$ is close to the time
it takes a fermion to traverse half the chain. At $T_{max}$ a
fermion initially, say, in the left half of the chain would have the
peak of its left wavepacket reflected at the left end and traveled a
distance $d$ from that end, while the peak of its right wavepacket,
yet to encounter an end, would be about the same distance $d$ from
the right end.  This implies that for each fermion, the amplitudes
of its presence at the two ends will be nearly equal at $T_{max}$.
$\Delta_2\sim 0$ (free model) can ensure an independent dynamics of
each fermion. Both $\ket{\mathcal{N}_1}$ and $\ket{\mathcal{N}_2}$
ensure that each fermion crosses either an even or an odd number of
others to reach either end, so that there is no net effect of
exchange phases. Thus each fermion will independently contribute a
state very close to $|\psi^{+}\rangle$ to $\rho_{1,N}(T_{max})$ (the
nearer to an end a fermion is initially, the more its contribution
deviates from $|\psi^{+}\rangle$). The cumulative contribution of
the {\em same} entangled state from all fermions at $T_{max}$
creates a very high entanglement between $1$ and $N$. If the
different fermions contributed distinct entangled states (e.g.
$\ket{\downarrow, \uparrow}+e^{i\phi}\ket{\uparrow, \downarrow}$
with different $\phi$'s), the cumulative entanglement could have
been vanishingly small even if block entanglement would be high.

  We now provide an estimate for the supply of impure pairs required as input in order to produce one almost pure maximally
entangled state, i.e. $f^{\text{out}}\geq 0.99$. At time
$T_{\text{max}}$, our mixed state becomes $\rho_{1,N}\simeq
f\ket{\psi^+}\bra{\psi^+}+\frac{(1-f)}{2}(\ket{\uparrow,
\uparrow}\bra{\uparrow, \uparrow}+\ket{\downarrow,
\downarrow}\bra{\downarrow, \downarrow})$, and the so called
recurrence method of purification, described in detail in
\cite{Bennett1996}, simplifies decisively (error correction, as
usable for communication through disordered chains
\cite{Allcock2008}, could also be used). Starting from an ensemble
of impure pairs with individual $f=0.544$, which is the value for
$N=151$ in Fig.~\ref{fig.2}, will require $5$ iterations of the
purification scheme on $\sim 361$ input pairs to achieve a single
pair with $f^{\text{out}}= 0.996$. In comparison, for a particular
chain of $N=9$ spins, which is also a representative number for
possible experiments, we have initially $f=0.9117$ and will need
$\sim 3$ impure pairs to be purified into an almost perfect
$\ket{\psi^{+}}$ in a single iteration step until our threshold is
exceeded by $f^{\text{out}}= 0.991$. We have also studied
numerically and plotted in Fig.\ref{fig.2} more general quenches.
The results suggest that rather than having a perfect N\' eel order
in the initial state ($\Delta_1\rightarrow\infty$), it is more vital
to propagate according to the free fermion Hamiltonian
($\Delta_2=0$) in order to achieve purifiable entanglement for
larger $N$. Our scheme is also suprisingly robust towards the type of disorder under consideration, i.e. random couplings. As shown in the left inset of Fig.\ref{fig.2}, for random offsets as large as an average ten per cent of the coupling strength $J$ the amplitude of the fully entangled fraction at $T_{max}$ as well as $T_{max}$ itself remain virtually unaffected.


The verifiability of our results is within experimental reach.
Promising realizations of $XXZ$ spin chains are Josephson junction
arrays \cite{Fazio2001}(where one can tune $J$ and $\Delta$ by
varying voltages \cite{Giuliano2005} or
 magnetic flux \cite{Lyakhov2005}). Spin-spin interactions in
optical lattices \cite{Duan2003} have recently been demonstrated
\cite{Trotzky2007}. These experiments also provide a global method
of preparing the initial state: pairwise spin triplets
$\ket{\psi^+}$ are first prepared in the double wells of a
superlattice, a magnetic field gradient is then applied to attain a
$\ket{\uparrow, \downarrow}$ state in each double well, and finally,
the long lattice instantaneously ramped down to create a N\'eel
state. Ensembles of finite lattices should be simulable with
superlattices and measurements of spin states are possible
\cite{Trotzky2007}.  We have ignored decoherence for the moment in
view of the relevant realizations (optical lattices are an ideal
arena to study quenches in absence of environments
\cite{Cramer2008b}, and small $N$ Josephson junction arrays do not
have significant decoherence over our ($T_{\text{max}}<\frac{N}{J}$)
time-scales \cite{Lyakhov2005}). Trapped ions are also known to
simulate spin chain Hamiltonians and their non-adiabatic changes
\cite{Porras2004}. In future, it would be interesting to study the
influence of a finite quench rate on the entanglement between the
end spins. We started from a Hamiltonian whose eigenstates are
product (un-entangled) states when written in terms of a local basis
{\em i.e.,} the spin states at given sites of a lattice. We then
quenched to a nearly free Hamiltonian which has momenta-like
eigenstates. Thus a similar behavior may show in other models, e.g.
Hubbard Hamiltonians when one quenches from an appropriate product
of localized states to a free (such as the superfluid) parameter
regime. Subsequent to the the work reported here, similar
entanglement dynamics has also been found by other authors
\cite{Galve2008} in driven ferromagnetic chains, which maps to our
case in certain limits.

{\em Summary:-} Non-equilibrium dynamics of a spin chain after a
quench
\cite{Sengupta2004,Cramer2008,Calabrese2005,DeChiara2006,Cincio2007}
is an area of intense activity. The scaling of distant spin-spin
correlations after quenches is topical
\cite{Igloi2000,Calabrese2007} with interesting instances of
polynomial scaling \cite{Igloi2000}. Much harder is to make the
truly ``quantum" correlations (or entanglement) between individual
distant spins ``substantial" and scale polynomially, which we have
shown to be possible through a certain global quench. This is
important because the entanglement between individual spins in
complex systems, a quantity of wide interest \cite{review2008}, is
easily measurable through correlation functions, but is notoriously
short ranged \cite{Osborne2002}. Our scaling of entanglement ($\sim
N^{-0.22}$) seems significantly superior to that in the statics or
dynamics of any spin model studied so far (except for specialized
couplings \cite{CamposVenuti2006}). Additionally, that we start from
a mixed state and obtain so much entanglement between distant spins
that it is efficiently purifiable to $\ket{\psi^{+}}$, at least for
small ($N\sim 9$) chains, and usable for linking solid-state based
quantum registers is interesting. Tests of our results, e.g., in
finite chains in optical superlattices \cite{Trotzky2007}, seem
feasible.

\begin{acknowledgments}
The studentship of HW is supported by the EPSRC, UK. SB acknowledges
EPSRC grant EP/D073421/1, the QIPIRC, the Royal Society and the
Wolfson Foundation. We thank T. Boness, P.Sodano, A. Kay and
particularly M. Cramer for very helpful comments.
\end{acknowledgments}

\end{document}